\documentclass[12pt,british]{iopart}
\usepackage[T1]{fontenc}
\usepackage[latin9]{inputenc}
\usepackage{babel}
\usepackage{amstext}
\usepackage{graphicx}
\usepackage[unicode=true,
 bookmarks=true,bookmarksnumbered=false,bookmarksopen=false,
 breaklinks=false,pdfborder={0 0 1},backref=false,colorlinks=false]
 {hyperref}
\hypersetup{pdftitle={Positive streamer inception and propagation in high-purity nitrogen: effects of voltage rise-rate},
 pdfauthor={S. Nijdam}}
\usepackage{breakurl}

\makeatletter
\usepackage{iopams}
\usepackage{setstack}

\usepackage{babel}

\@ifundefined{showcaptionsetup}{}{%
 \PassOptionsToPackage{caption=false}{subfig}}
\usepackage{subfig}
\makeatother

\begin{document}

\title[Voltage rise-rate effects on positive streamers in pure N$_{2}$]{Inception and propagation of positive streamers in high-purity nitrogen:
effects of the voltage rise-rate}

\author{T.T.J. Clevis$^{1}$, S~Nijdam$^{1}$ and U~Ebert$^{1,2}$}

\address{$^{1}$ Eindhoven University of Technology, Dept.\ Applied Physics\\
 P.O. Box 513, 5600 MB Eindhoven, The Netherlands}

\address{$^{2}$ Centrum Wiskunde \& Informatica (CWI), Amsterdam, The Netherlands}

\ead{s.nijdam@tue.nl}
\begin{abstract}
Controlling streamer morphology is important for numerous applications.
Up to now, the effect of the voltage rise rate was only studied across
a wide range. Here we show that even \emph{slight} variations in the
voltage rise can have significant effects. 

We have studied positive streamer discharges in a 16\,cm point-plane
gap in high-purity nitrogen 6.0, created by 25\,kV pulses with a
duration of 130$\,$ns. The voltage rise varies by a rise rate from
1.9$\,$kV/ns to 2.7$\,$kV/ns and by the first peak voltage of 22
to 28$\,$kV. A structural link is found between \emph{smaller} \emph{discharges}
with a \emph{larger inception cloud }caused by a \emph{faster rising
}voltage. This relation is explained by the greater stability of the
inception cloud due to a faster voltage rise, causing a delay in the
destabilisation. Time-resolved measurements show that the inception
cloud propagates slower than an earlier destabilised, more filamentary
discharge. This explains that the discharge with a faster rising voltage
pulse ends up to be shorter.

Furthermore, the effect of remaining background ionization in a pulse
sequence has been studied, showing that channel thickness and branching
rate are \emph{locally }affected, depending on the covered volume
of the previous discharge.
\end{abstract}

\submitto{\JPD}

\maketitle

\section{Introduction\label{sec:BG-ionization-Introduction}}

Streamers occur in nature in the streamer corona of lightning leaders
in, between and below thunderclouds, and in sprites, jets and other
transient luminous events above thunderclouds. Streamers have numerous
applications in gas and water cleaning~\cite{Clements1989,Grabowski2006,Veldhuizen2000,Winands2006a},
ozone generation~\cite{Veldhuizen2000}, particle charging~\cite{Kogelschatz2004,Veldhuizen2000}
and flow control~\cite{Moreau2007,Starikovskii2008}. They appear
in a large variety of gases, including air, syngas, industrial exhaust
gases, pure nitrogen, oxygen, argon and carbon dioxide~\cite{Nijdam2010,NijdamThesis}
and the atmospheres of other planets~\cite{Dubrovin2010}. Diameters
and velocities of streamers in air can vary by orders of magnitude~\cite{Briels2006,Briels2008};
this has direct consequences for the efficiency of the conversion
of electric power into chemical products~\cite{Winands2006,Heesch2008}.
Thick, fast and efficient streamers appear when the voltage rises
to a sufficiently high value within a sufficiently short rise time,
of the order of several to tens of nanoseconds~\cite{Briels2008,Starikovskii2008,Yagi2011,Winands2008a}.
However, up to now it has not been clearly understood how different
voltage rise times bring different streamer structures about. Here
we present results that suggest a significant influence of slight
variations of the voltage rise time on the discharge characteristics.
Also, a qualitative comparison with theory is given.

Here we present experimental results regarding the influence of the
voltage pulses' rising edge on streamer channel formation and propagation
of positive streamer discharges. Research on the influence of the
voltage rise on discharges has been done by for instance Briels \emph{et
al.}~\cite{Briels2006}, Yagi \emph{et al.}~\cite{Yagi2011} and
Zhang \emph{et al.~}\cite{Zhang2008}. All these experiments were
performed across a wide parameter range. For instance, Briels \emph{et
al. }have used rise times between $23\,$ns and $60\,$ns at a wide
range of DC-voltages ($11.3\,$kV - $60\,$kV) and pulse lengths ($25\,$ns-120$\,$ns).\emph{
}Yagi uses $15\,$kV voltage pulses with rise times between $7\,$ns
and $40\,$ns at atmospheric pressure. Finally, Zhang \emph{et al.}
apply $20\,$kV pulses with $20\,$ns and $200\,$ns rise time at
$5-20\,$kPa. Also the configurations vary from point-plane geometry
(Briels \emph{et al.}) to coaxial discharges (Yagi \emph{et al.})
and line-plane (Zhang \emph{et al.}). 

Because of this large parameter range, a difference in their conclusions
is expected. Yagi \emph{et al.} find that a \emph{faster} voltage
rise leads to \emph{faster} streamers. Briels \emph{et al. }report
the opposite, namely that for a faster voltage rise, the streamers
are \emph{slower.} This they explain by their larger diameter, reducing
the electric field enhancement. Zhang \emph{et al.} do not measure
propagation velocity, but also observe \emph{thicker} channels for
shorter rise time. They attribute this to a higher number of individual
avalanches that overlap.

In none of the mentioned researches, the effect of \emph{slight }variations
in the voltage rise has been investigated. This will be done here.
We use pure nitrogen to minimize the effect of photo-ionization on
streamer propagation~\cite{Nijdam2010}. This also reduces the size
of the inception cloud and therefore makes it easier to study streamer
channel formation in the pressure range we use (around 100\,mbar).
We have chosen this pressure range because in this range the streamers
are almost or barely crossing the 16\,cm in our set-up gap during
our pulse duration of 130 ns.

\section{Experimental setup}

All experiments presented here were performed in a point-plane electrode
configuration with a separation of $160\,$mm, enclosed in a roughly
$50\times50\,$cm vacuum vessel. This geometry provides high electric
fields around the point. As a result, the discharge starts near this
point and propagates towards the plane. The discharges were created
in 100\,mbar nitrogen 6.0, which is high purity nitrogen with less
than $1\,$ppm total impurities. To minimise additional impurities
in the vessel, it is pumped down to about $10^{-7}\,$mbar outside
measurement periods and flushed three times with nitrogen 6.0 before
use. Besides, a constant gas flow of $400\,$sccm through the vessel
is used during measurements, resulting in a fresh gas fill every twenty-five
minutes. 

\begin{figure}[t]
\centering \includegraphics[width=10cm]{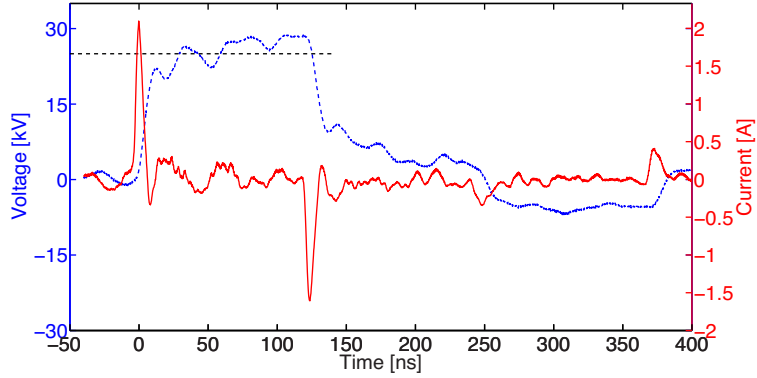} \caption{Averaged waveforms of thirty 25\,kV Blumlein pulses under vacuum
condition (i.e. without discharge). The voltage is shown as the dashed
(blue) line. The continuous (red) line shows the current waveform,
dominated by the displacement current C dV/dt. The horizontal dashed
(black) line indicates the average voltage of the pulse as we report
it.}

\label{fig:wf-average} 
\end{figure}

The high voltage pulse is produced by a Blumlein pulse forming network,
creating $15\,$--$35\,$kV pulses with $130\,$ns duration and $10\,$ns
rise and fall time. All measurements in this article have been done
with positive 25\,kV pulses, creating positive streamers emerging
from the point electrode. A typical 25\,kV voltage pulse and the
corresponding current under vacuum conditions are shown in figure~\ref{fig:wf-average}.
The voltage is measured with a Northstar PVM-4 1:1000 voltage probe
with a bandwidth of $110\,$MHz. The current is determined from the
voltage across a $50\,\Omega$ shunt resistor between the cathode
plane and the grounded vessel. Because the current in figure~\ref{fig:wf-average}
is measured under vacuum conditions, it is purely capacitive:$C\cdot dV/dt$.
The measurements indicate that the vessel capacitance is $C\approx1\,$pF.

The streamer discharges are imaged with a Stanford Computer Optics
4QuickE iCCD-camera and a Nikkor-UV $105\,$mm $f/4.5$ UV-lens. This
camera is capable of nanosecond delay and shutter control, allowing
for precise image timing. More details on the setup can be found in~\cite{Nijdam2010}.

\section{Results}

\subsection{The variations of the voltage pulse}

\begin{figure}[t]
\centering\includegraphics[width=10cm]{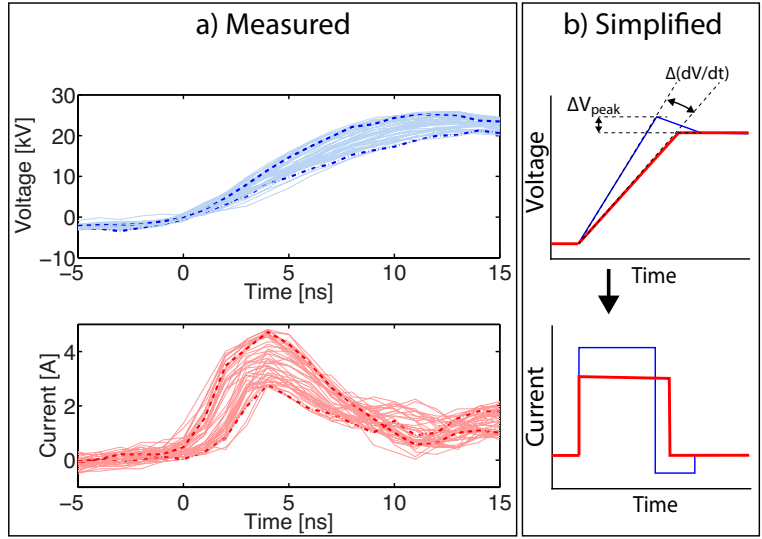} \caption{\textbf{a)} Sequence of measured voltage (above) and current (below)
waveforms zoomed in time into the rising edge of the voltage pulse.
Voltage slope and peak voltage vary between different pulses, which
also causes a variation of the height and width of the current peak.
To emphasize the variation, two of the outer waveforms are plotted
with bold dashed lines in both the current and the voltage graph.
\textbf{b)} Upper panel: Simplified scheme of the voltage variation
defining it as a variation in the slope of the voltage and the peak
voltage. Lower panel: Simplified scheme of the corresponding displacement
current measured through the cathode plate.}

\label{fig:wf-zoom} 
\end{figure}

The voltage waveforms measured in our setup vary slightly from pulse
to pulse in the rising and falling edges of the voltage pulses. These
variations are related to the pulse forming network and we can (currently)
not control it. The upper panel in figure~\ref{fig:wf-zoom}a shows
a series of measured voltage waveforms as a function of time during
the time interval from $t=-5\,$ns up to $t=15\,$ns around the rising
edge. The time axis is defined as in figure~\ref{fig:wf-average},
with the voltage pulse starting to rise at $t=0\,$ns. The lower panel
in figure~\ref{fig:wf-zoom}a shows the corresponding current waveforms.
The waveforms shown here are measured under discharge conditions but
are essentially identical to waveforms under vacuum conditions because
the corona current is still negligible at the start of the voltage
pulse.

The variation in the voltage wave form can be described in a simplified
manner in terms of \emph{voltage slope} and \emph{peak} \emph{voltage}.
Figure~\ref{fig:wf-zoom}b illustrates this simplified sketch for
the voltage as well as for the corresponding capacitive current. Though
the transitions in the actually measured waveforms are, of course,
much smoother, the simplified schematic is useful to describe the
variation.

For faster rising voltage pulses (with larger voltage slope $\textrm{d}V/\textrm{d}t$),
the peak voltage $V_{\textrm{peak}}$ is larger as well, as it shows
 a short voltage overshoot. Taking $\textrm{d}V/\textrm{d}t$ between
$t=2\,$ns and $t=10\,$ns as a measure, we find that the average
slope varies between $1.9\,$kV/ns and $2.7\,$kV/ns in the measured
waveforms. The maximal difference of the peak voltage $\Delta V_{\textrm{peak}}$
at $t=12\,$ns is $6\,$kV.

As far as we have been able to test, the voltage slope varies in a
completely random manner. It is neither alternating, nor remaining
for a while in one state and then jumping to another state. Our hypothesis
for the physical mechanism behind the voltage rise variation is based
on the breakdown of our multiple spark-gap. As a part of the Blumlein
pulse forming network, the multiple spark-gap consists of seven electrodes
that distribute the voltage over six gaps. After the trigger pulse
the gaps break down one by one, leaving the full voltage distributed
over less many gaps after every gap breakdown. At the final breakdown,
the last gap releases the full DC-voltage.

We think that the \emph{order} of breakdown of the individual gaps
may cause the voltage rise variation. The trigger voltage is put across
the first and second electrode, thereby changing not only the voltage
over the first gap voltage, but also over the second gap. As a consequence,
it is not a consecutive breakdown of adjacent gaps that evolves in
the multiple spark-gap, but an interplay between gap voltage, gap
distance and stochastic variations that lead to variable breakdown
sequences and delays. For the multiple spark-gap as a whole, this
means that the total breakdown delay may vary, resulting in slightly
different voltage rise rates.

\subsection{Discharge observations}

\begin{figure}[t]
\centering \includegraphics[width=12cm]{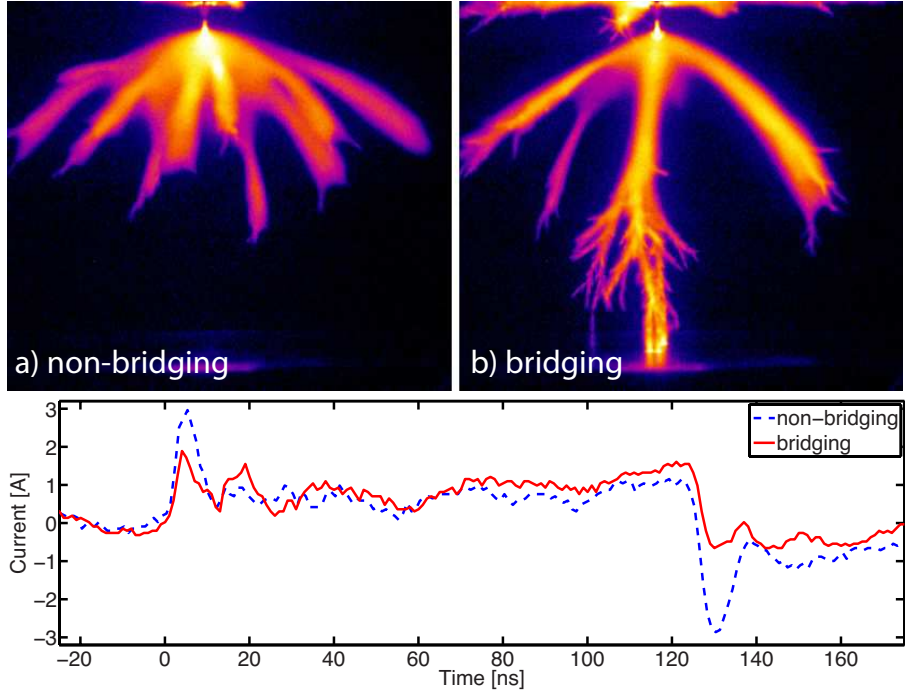} \caption{Comparison of \textbf{(a)} bridging and \textbf{(b)} non-bridging
streamer discharges and of their current waveforms (lower panel).
At the start of the discharge, up to 10\,ns, the non-bridging discharge
has a significantly higher first current peak. After about 125\,ns
the current of the non-bridging discharge drops to about -3\,A, while
the current of the bridging discharge drops only to just below zero.
We will relate these differences to the voltage wave form and to the
corona current. }

\label{fig:comparison} 
\end{figure}

We have observed a structural variation of the discharge behaviour
that is related to, and most likely caused by, the voltage pulse variation.
Panels a and b in figure~\ref{fig:comparison} show two discharges,
and the lower panel shows the corresponding current waveforms. These
discharges occurred in the same pulse sequence with fixed parameters:
100\,mbar (nitrogen 6.0), 25\,kV Blumlein pulses at 5\,Hz repetition
rate. Note that it has been shown  previously~\cite{Nijdam2011c}
that at this repetition rate the influence of leftover background
ionization from the previous discharge pulse is significant, creating
thick and smooth channels rather than thin and feather-like branches
like at low repetition rates. Other interesting aspects of the discharge
morphology will be discussed later in section~\ref{sec:Discharge-morphology:-local}.

The discharge in figure~\ref{fig:comparison}a does not reach the
cathode, while the discharge in figure~\ref{fig:comparison}b clearly
does. In the discharge sequence of this measurement, the ratio of
\emph{bridging to non-bridging} discharges was about 50/50. Though
the difference in the size of the two discharges is large, one must
realise that during the latest stage of evolution, the high electric
field between the approaching streamer and the cathode plane in the
bottom part of the gap as well as possibly secondary emission from
the cathode cause a significant acceleration of the channel propagation.

\begin{figure}[!t]
\centering \includegraphics[width=14cm]{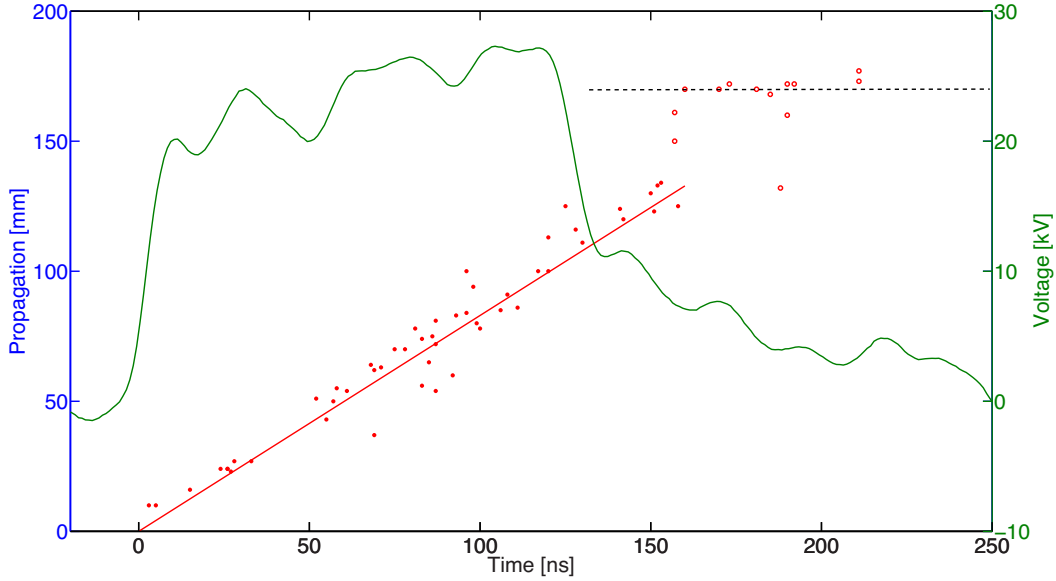} \caption{Red dots: Length of the longest streamer channel as a function of
image exposure time. The linear fit (solid red line) indicates a streamer
velocity of $(8.3\pm0.5)\cdot10^{5}\,$m/s. The solid green line is
the voltage. The black dashed line at 170$\,$mm indicates the cathode
plane. At a distance of \textasciitilde{}40 mm from the cathode plane
the streamers largely accelerate  towards it.}

\label{fig:velocity} 
\end{figure}

Figure~\ref{fig:velocity} shows the length of the longest streamer
channel as a function of time. The longest streamer channel of a discharge
has been taken as a measure for the propagation, because it gives
a lower boundary for the maximum velocity. Due to the 2D-projection
of the discharge on the camera CCD, streamer channels with a component
perpendicular to the image plane will appear shorter on the image.
Therefore the longest streamer is the best measure for the actual
streamer length. Note that all data points represent different discharges,
because the camera cannot capture multiple images within one discharge.
As a reference, a voltage pulse is also plotted in the figure. The
start of the streamer propagation has been aligned with the rise of
the voltage pulse.

The figure demonstrates that up to $t=150$\,ns the streamers propagate
with nearly constant velocity. The linear fit indicated by the solid
red line corresponds to a velocity of $(8.3\pm0.5)\cdot10^{5}\,$m/s.
After $t=150$\,ns, or at a distance of \textasciitilde{}40\,mm
from the cathode plane, the streamers seem to `jump' to the cathode
plane whose position is marked with a dashed black line. Streamer
lengths near this line represent discharges that have bridged the
gap. The final streamer acceleration over the last 40\,mm even after
the main voltage pulse is probably caused by secondary electron emission
from the cathode plane, combined with the field enhancement due to
the streamer head charge approaching the planar electrode. In this
discharge series, only one discharge has not bridged the gap after
$t=150$\,ns (data point at $t=185$\,ns and $x=130$\,mm). As
stated before, the bridging to non-bridging ratio of the discharge
series discussed in the rest of this article is about 50/50. The fast
jump towards the cathode plane over the last 40 mm indicates that
the difference in terms of propagation between the two discharges
in figure~\ref{fig:comparison} is quite small. Whether a discharges
does or does \emph{not} bridge, could thus in principle be caused
by stochastic variations.

However, distinguishing between bridging and non-bridging discharges
and analysing the corresponding current waveforms brings up structural
differences related to the voltage rise variation. In figure~\ref{fig:comparison},
these current waveforms are shown below the discharge images. At the
end of the voltage pulse, around $t=125$\,ns, there is a significant
difference in the current between the bridging and non-bridging discharges.
While the \emph{non-bridging} has a strong negative peak of about$-3\,$A
that corresponds to the displacement current $C\cdot dV/dt$, the
current of the \emph{bridging} discharge decays only to just below
zero. This can be attributed to the higher corona current associated
with the bridging of the gap. Because the conductive connection with
the cathode plane is formed just before the pulse voltage decreases,
a significantly higher corona current flows as opposed to the non-bridging
discharge. The negative capacitive current due to the voltage drop,
which is clearly visible for the non-bridging discharge, is neutralised
by the corona current.

The current waveforms in figure~\ref{fig:comparison} also show a
significant difference at the early stage, in the height of the first
peak at $t\approx6\,$ns. As explained in the previous subsection,
this peak corresponds to the capacitive current of the voltage pulse
rise. In a measurement series of about thirty discharges, we have
observed that \emph{structurally} the \emph{bridging} discharge has
a \emph{lower} first current peak. This makes it plausible that the
bridging behaviour is linked to the voltage pulse variation.

\subsection{Discharge behaviour related to the voltage pulse}

\begin{figure}[t]
\centering \includegraphics[bb=0bp 0bp 966bp 705bp,clip,width=10cm]{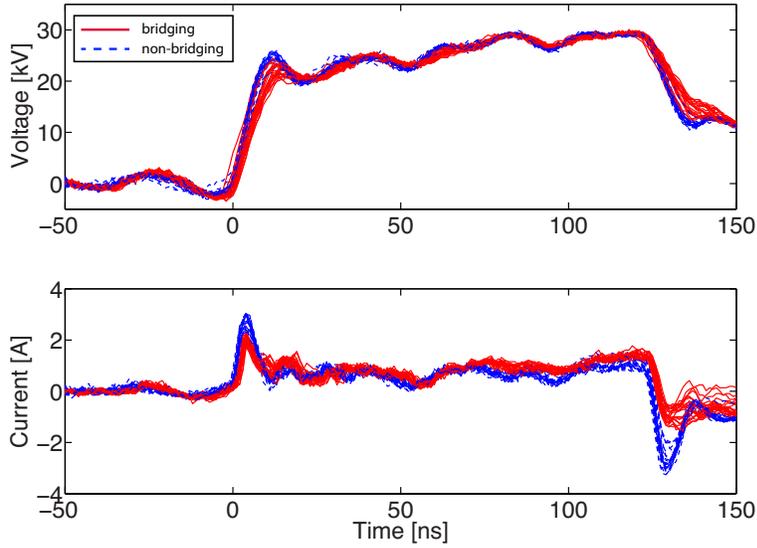}
\caption{Matched \emph{individual} waveforms of voltage (upper panel) and current
(lower panel) in a discharge sequence. The bridging discharges are
indicated in red, the non-bridging ones in blue. The bridging discharges
have no negative current peak at the end of the voltage pulse. The
figure shows that the first current peak of the bridging discharges
is always lower than for the non-bridging discharges. According to
the discussion in the previous subsection, the bridging discharges
are therefore created by voltage pulses with a lower voltage rise
time.}

\label{fig:pulsevariation} 
\end{figure}

Figure~\ref{fig:pulsevariation} shows the individual voltage and
current waveforms of the full discharge sequence. The solid red lines
mark all discharges that clearly do \emph{not} have the strongly negative
current peak at the end of the voltage pulse. As discussed in the
previous section, this means that these discharges have bridged the
gap. The dashed blue lines, on the other hand, indicate the \emph{non-bridging}
discharges.

Just after the start of the discharge pulse, the first current peak
of the bridging discharges is clearly lower than for the non-bridging
ones. The average peak height is $2.0\pm0.2\,$A for the bridging
discharges versus $2.8\pm0.3\,$A for the non-bridging discharges,
which amounts to a significant difference of about a $3\sigma$. The
upper panel in figure~\ref{fig:pulsevariation} shows that the lower
initial current peak of the bridging discharge indeed corresponds
to a lower voltage rise time of the voltage switch, as argued earlier
based on the capacitive nature of this initial current.

\section{Physical interpretation}

\subsection{Formation and size of the inception cloud}

How can it be explained that a faster rising and overall higher voltage
creates  discharges that propagate less far within the same time?
Figure~\ref{fig:comparison} shows another difference between bridging
and non-bridging discharges: the size of the inception cloud. As shown
previously in~\cite{Briels2008c,Nijdam2010,Nijdam2011c,Nijdam2011,Ebert2011},
next to a needle electrode, first an inception cloud grows. This cloud
forms an ionization front that propagates outward and is visible as
a shell. The shell eventually destabilizes into streamers. Analysing
the discharge series, it seems that a \emph{faster} \emph{voltage
rise} causes a \emph{larger} \emph{inception} \emph{cloud} and correspondingly
a \emph{later} emergence of streamer channels. 

The maximal radius $R_{max}$ of the inception cloud as a function
of breakdown field $E_{\textrm{br}}$ and applied voltage V can be
estimated as $R_{max}=V/E_{\textrm{br}}$. This estimate is based
on the assumption that the inception cloud is spherical and perfectly
conducting; then if the cloud is on an electric potential V, and if
the electric field on the surface is precisely the breakdown field,
the radius is given by the equation above. We remark that the volume
of the inception cloud is actually larger than the critical volume
(where the electric field exceeds the breakdown field) in the absence
of the discharge. 

The breakdown field $E_{\textrm{br}}$ in 100\,mbar nitrogen is approximately
$2\,$kV/cm. The maximal radius of the inception cloud is therefore
125\,mm for a voltage of 25\,kV; this is comparable to the size
of the gap. However, even in air with its stabilizing effect of photo-ionization,
the maximal radius is typically not reached in positive discharges
(though a counter example is shown in figure~\ref{fig:oxygen_sander}d
below) while for negative discharges an example can be found in~\cite{Nijdam2011}
where the discharge cloud reaches approximately the maximal radius.
The important fact is, however, that a faster voltage rise creates
a more ionized and more stable inception cloud that breaks up \emph{later}
into individual streamer channels.

\subsection{The sizes of the whole discharge and of the inception cloud}

\begin{figure}[t]
\centering\includegraphics[width=8cm]{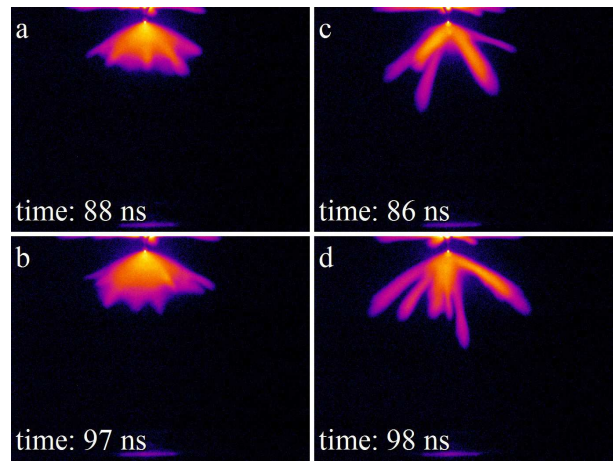} \caption{Time resolved measurements of the discharges. The gate closing time
relative to the start of the discharge is indicated in each frame.
Images (a) and (b) show a larger inception cloud and these discharges
are smaller as a whole. In images (c) and (d) both discharges have
a smaller inception cloud and have propagated further within the same
time.}

\label{fig:timeresolved}
\end{figure}

Now the evolution of the whole discharge will be related to the size
of the inception cloud. We took time-resolved images of the discharge
propagation. Figure~\ref{fig:timeresolved} shows two sets of images,
the upper ones with an exposure time of about 87\,ns, the lower ones
with about 97\,ns. In both cases, the discharges on the left have
much larger inception clouds and a smaller overall extension, while
on the right the inception clouds are small and the emerged streamers
have propagated much further out. We conclude that the inception cloud
propagates slower than the individual channels. The explanation lies
in the fact that at the same distance from the electrode, the electric
field at the tip of a pointed streamer channel is enhanced to much
higher values than at the rather flat ionization front around the
inception cloud. The higher local field leads to a higher velocity. 

Note that figures~\ref{fig:timeresolved}a and b show, that the inception
cloud destabilises about $70-90\,$ns after the start of the voltage
pulse. As the rise time of the voltage pulse is only about $10\,$ns,
the actual differences in the inception cloud size are still visible
much later, long after the voltage rise. We hypothesize that the more
strongly driven early nucleation process of the discharge near the
electrode needle during the first 10 ns still influences the stability
of the cloud after 60 - 80 ns. The underlying reason could be the
higher conductivity of the plasma directly at the electrode which
is created at the early stages; a high conductivity there allows a
continuous flow of electric current from the electrode into the plasma
region. Simulations investigating this hypothesis are currently under
development. Note that once the plasma front becomes unstable and
starts to branch, the process cannot be reversed. It may continue
in that form or branch even further, but it will never return to the
homogeneous plasma front it originated from.

\begin{figure}[t]
\centering \includegraphics[width=14cm]{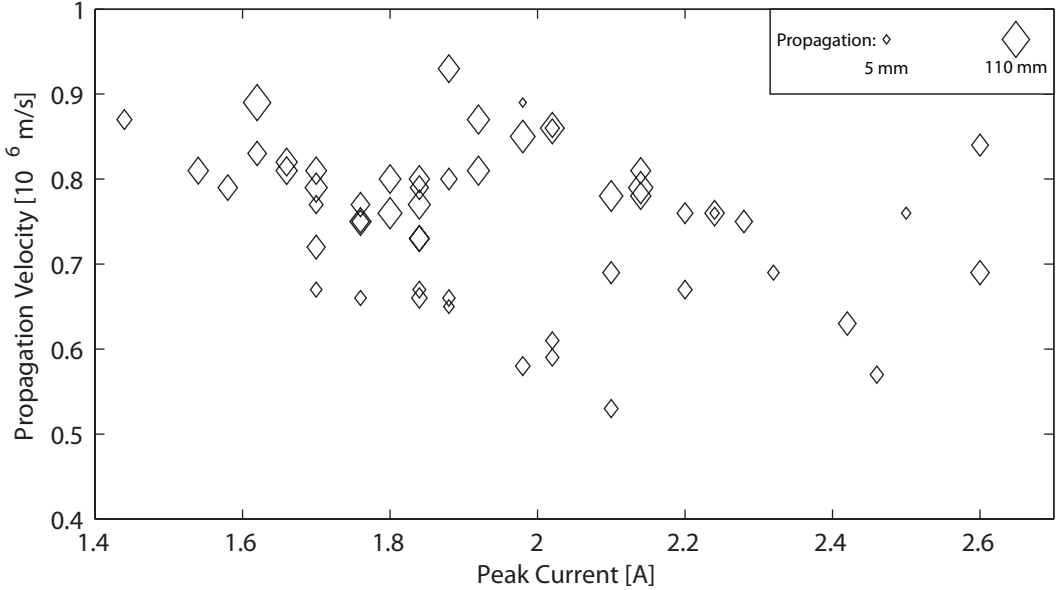}

\caption{Mean propagation velocity as a function of the peak current. We recall
that a higher peak current corresponds to a shorter voltage rise time.
The size of the data points indicates the propagation distance. The
velocity is determined by dividing the propagation length through
the end-time of the exposure (relative to the start of the voltage
pulse) and therefore represents the average velocity between the start
of the pulse and the end of the exposure. }

\label{fig:timing_graph} 
\end{figure}

It appears that the inception cloud propagates slower than the streamers
at the same distance from the electrode. This behaviour is further
investigated in figure~\ref{fig:timing_graph}. It shows the average
propagation velocity of the discharge during the exposure time as
a function of the peak current (which is proportional to the voltage
rise rate). The propagation distance is indicated by the size of the
markers. Small diamonds represent measurements early in the discharge
formation, whereas large diamonds correspond to almost fully developed
discharges (discharges up to $110\,$mm length have been measured).
The larger diamonds show that the final average velocity is about
$8\cdot10^{5}\,$m/s. For higher peak currents, there are relatively
many data points for low velocities. These are mostly small diamonds,
showing that the discharge is still small. The figure shows that indeed
the discharge propagates more slowly when the peak current is higher
(and thus the voltage rise faster).

Another aspect that supports our observation can be found in figure~\ref{fig:velocity}.
Between 60 \,ns and 100 \,ns, a couple of data points significantly
deviate from the trend through a shorter propagation length. From
the corresponding images we have found that these are discharges with
a larger inception cloud, from which no channels have emerged yet
- similar to those on the left in figure~\ref{fig:timeresolved}.

We conclude that higher voltage rise times create inception clouds
that grow to a larger size before they destabilize into streamers,
and that the inception cloud grows more slowly than a streamer at
the same distance from the electrode. This is very reasonable, since
the electric field at the edge of an inception cloud is lower than
at a streamer tip at the same distance from the electrode.

\subsection{The inception cloud in air}

\begin{figure}[t]
\centering \includegraphics[width=1\textwidth]{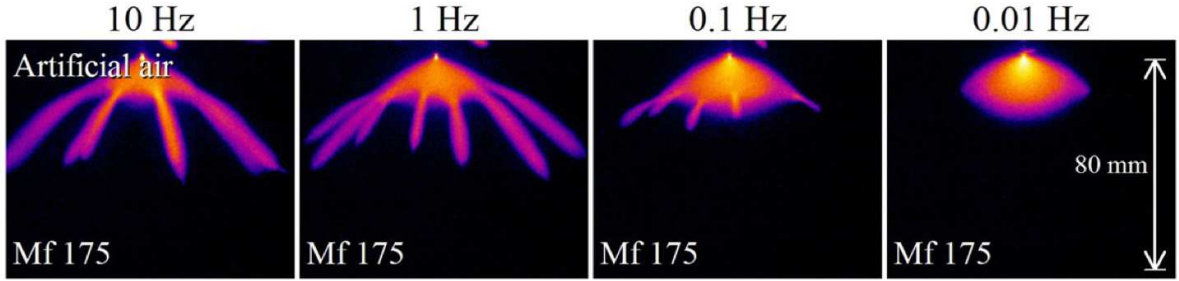} \caption{Part of figure 3 from\emph{~}\cite{Nijdam2011c}\emph{. }The discharges
are created with the same electrodes and voltage pulses as in the
present paper, but in artificial air at 200$\,$mbar (rather than
in nitrogen 6.0 at 100 mbar), and at different pulse repetition rates
as indicated. The same relation is found as in figure~\ref{fig:timeresolved}:
A larger inception cloud corresponds to a smaller overall discharge
size. }

\label{fig:oxygen_sander} 
\end{figure}

The size of the inception cloud and of the overall discharge behaves
similarly in air; as can be found in an investigation of the effects
of background ionization on streamer discharges in air and in nitrogen~\cite{Nijdam2011c}.
Figure 3 in that paper compares the observed effects in air and nitrogen.
The paper as well as our figure~\ref{fig:oxygen_sander} that is
reproduced from~\cite{Nijdam2011c} show that discharges in air behave
similarly as discharges in nitrogen 6.0: a larger inception cloud
coincides with a smaller overall discharge size within the same evolution
time. Based also on our theoretical understanding, we hypothesize
that this behaviour is valid in other gases as well. We also expect
the influence of the voltage rise rate to be a general discharge effect,
independently of the gas composition.

We conclude with two remarks on the discharge in air shown in figure~\ref{fig:oxygen_sander}.
1. The inception clouds in air are always much larger than in nitrogen
under similar conditions; this is due to the stabilizing effect of
the non-local photo-ionization reaction. The inception cloud in the
rightmost panel of the figure is almost 30 mm which is close to the
maximal radius predicted to be about 40 mm. This is the case even
though the discharge is positive, and therefore more unstable. 2.
Discharges at higher repetition rates develop a \emph{smaller} inception
cloud and their streamer channels propagate \emph{further}. The higher
repetition rate creates a higher background density~\cite{Nijdam2011c}.
But why a higher background ionization destabilizes the cloud earlier,
is not understood yet, as the traces of the previous discharges are
smoothed out by diffusion by the time of the next voltage pulse. In
nitrogen we see and understand the opposite effect.

\section{Discharge morphology: local branching reduction \label{sec:Discharge-morphology:-local}}

The discharges in figure~\ref{fig:comparison} have more interesting
features besides their relation to the voltage rise rate. It was already
noted by Nijdam \emph{et al.}\cite{Nijdam2011c} that at the present
repetition rate of 5\,Hz the remaining ionization of the previous
discharge pulse has a significant effect on the next discharge in
nitrogen 6.0. Nijdam \emph{et al. }conclude that the abundance of
seed electrons reduces the stochastic fluctuations of electron density
ahead of the ionization front of streamers or of the inception cloud,
and that they therefore destabilize and branch later than in virgin
gas~\cite{Luque2011a}. In the lower part of the discharge in figure~\ref{fig:comparison}b,
however, the channels are thinner and branch more, similarly to discharge
in a lower background ionization. This observation is consistent with
calculations in \cite{Nijdam2011c} that indicate that ionization
created in the upper part of the discharge will not reach the lower
discharge region in the time between consecutive voltage pulses.

An elegant experiment by Takahashi~\cite{Takahashi2011} shows that
streamer branching can be controlled locally by irradiating an (argon)
discharge at atmospheric pressure with a laser. From their results
they deduce that ionization levels higher than $5\times10^{5}\,$cm$^{-3}$
influence streamer branching. 

Both experiments show that the remaining ionization of a previous
discharge pulse has \emph{local} effects on the subsequent discharge
for the discharge volume and gas considered here.

\begin{figure}[t]
\centering\subfloat[]{\includegraphics[width=0.55\textwidth]{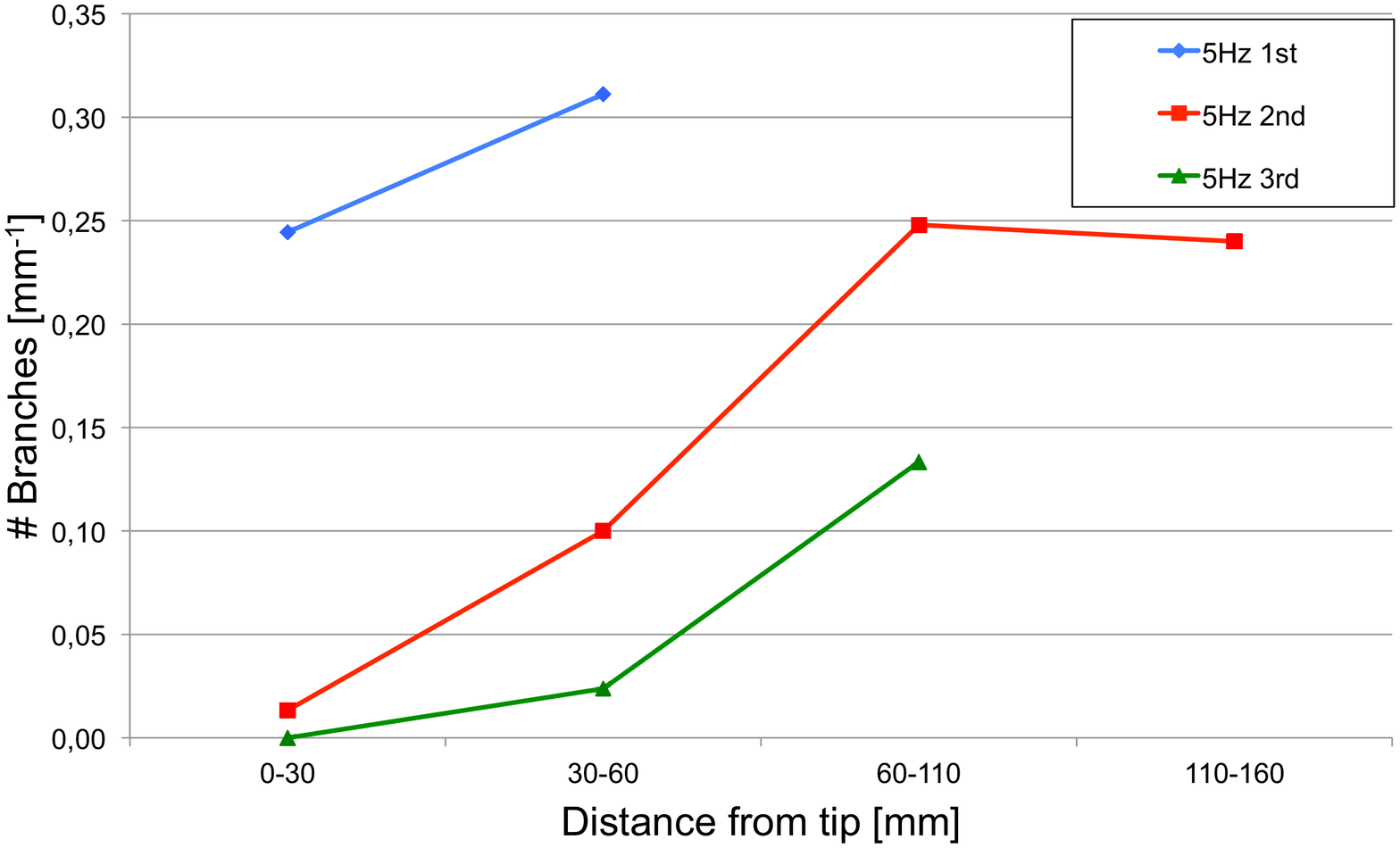}

}\subfloat[]{\includegraphics[width=0.4\textwidth]{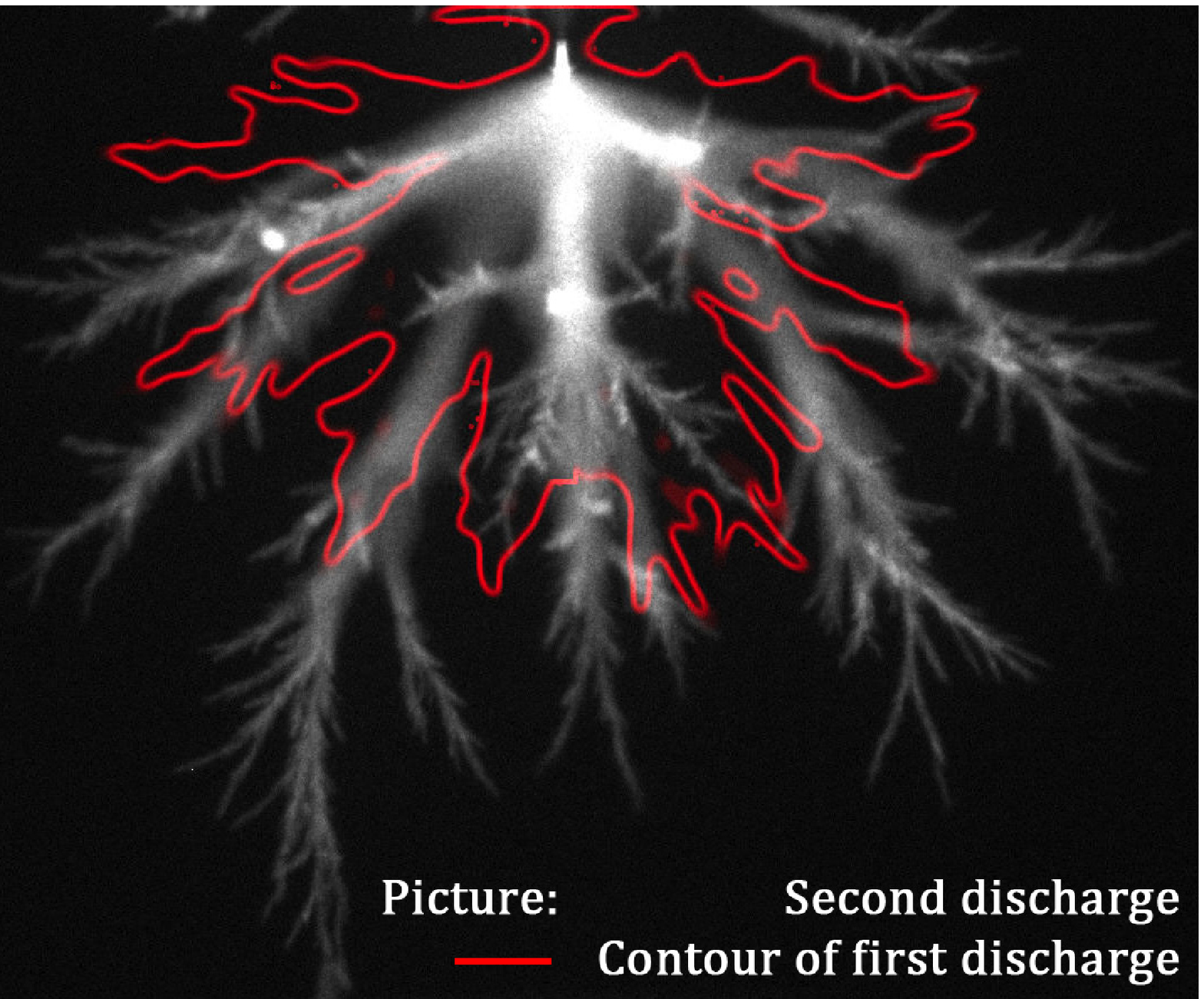}

} \caption{\textbf{a)} Measured number of `feathers' per mm streamer length for
three subsequent discharges at 5 Hz after a long pause. \textbf{b)}
Photograph of visualization of the second discharge with the red contour
of the first discharge overlaid. Conditions: 100\,mbar nitrogen 6.0,
25\,kV Blumlein pulses at 5\,Hz $\,$repetition rate.}

\label{fig:localbranching} 
\end{figure}

Similarly to the experiments performed by Takahashi, we have performed
measurements to verify that a higher background ionization density
reduces branching locally. The hypothesis is that at the next discharge
pulse, there is non-uniform ionization remaining from the previous
discharge(s). The pre-ionized volume and ionization density in pure
nitrogen is determined by diffusion and recombination. When diffusion
is fast, the remaining ionization is expected to be non-local and
uniform, whereas for slow diffusion we expect to see clear local effects
of the remaining ionization on the branching behaviour of the subsequent
discharge.

Figure \ref{fig:localbranching}a shows the branching rate of three
subsequent discharges after a long pause. The pause is much longer
than one minute, and experiments and calculations agree on the fact
that after more than one minute, there are no visible effects of previous
discharges left. The graph shows the number of (manually counted)
branches per millimetre of streamer channel length as a function of
the distance to the tip (grouped) for the first, second and third
discharge after the pause. The number of branches is evaluated by
selecting a certain part of a streamer channel and counting its branches.
It can be seen clearly that the branching rate decreases for subsequent
discharges, but only in the volume of the preceding discharge. 

The effect of the first discharge on the second one at 5\,Hz repetition
frequency is visualised in a different manner in figure~\ref{fig:localbranching}b;
here a picture of the second discharge is overlayed with a contour
sketch of the first discharge. Of course this is only a 2D-projection,
but it is clearly visible that branching is suppressed within the
contour of the first discharge. Note that the channels of subsequent
discharges do \emph{not }follow the exact same path (as was also shown
in~\cite{Nijdam2011c}). Branching is suppressed in the second discharge
within roughly the whole (semi-)spherical volume of the first discharge,
with a smooth transition outwards. This behaviour is consistent with
the estimated diffusion length of pre-ionization within the available
time, also calculated in~\cite{Nijdam2011c}, where the diffusion
length after 0.2\,s in 200\,mbar pure nitrogen was found to be about
7\,mm (see figure~6 of that work). At 100\,mbar this length will
double so that the pre-ionization is spread out over the whole (semi-)spherical
volume covered by the discharge.

To the best of our knowledge, this is the first experiment that shows
the \emph{build-up} of pre-ionization in the discharge volume in repetitive
discharges and its local influence on the morphology of subsequent
discharges.

Another peculiarity that can be observed in figures~\ref{fig:comparison}b
and~\ref{fig:localbranching}b: is the structure about two-thirds
of the distance between tip and plate. It shows a very sudden transition
from thick, non-branching channels to many much thinner and feather-like
branches that propagate into quite different directions. We name this
structure a `knotwilg', after the Dutch name for a \emph{pollard willow}
tree with a thick stem and many thin branches (that are harvested
regularly). We will address possible explanations for this structure
in future work.

\section{Conclusions}

We have investigated positive streamer discharges in high purity nitrogen
at 100\,mbar, generated by voltage pulses of 130\,ns duration at
5\,Hz repetition. We have shown that a slight variation in the voltage
rise rate between different pulses has an important impact on the
dynamics and morphology of the discharges. The voltage rise is characterised
by the slope, which varies from $1.9\,$kV/ns to $2.7\,$kV/ns, and
by the peak voltage, ranging from $22\,$kV to $28\,$kV.

Pulses with a \emph{faster} rise rate create more compact discharges
within the duration of the pulse and thus have a lower average propagation
velocity. This difference appears to arise during the first phase
of streamer formation, because the \emph{more compact} discharges
have a \emph{larger} inception cloud. Time resolved measurements indicate
that the inception cloud extends indeed slower than streamer channels.
This can be immediately understood as the electric field enhancement
at the edge of a (round) inception cloud is lower than at a (pointed)
streamer tip.

The maximal size of the inception cloud depends on the voltage. Approximating
the cloud as an ideally conductive sphere, its maximal radius is $R_{max}=V/E_{\textrm{br}}$,
where $V$ is the applied potential and $E_{\textrm{br}}$ the breakdown
field of the gas. Therefore both our understanding and our measurements
suggest that a faster voltage rise creates a larger and more stable
inception cloud, thereby delaying the destabilisation into streamer
channels. Therefore after the 130\,ns voltage pulse, the discharge
reaches out less far than with a lower voltage rise rate. We believe
that this relation between voltage rise rate, size of inception cloud
and spatial extension of the complete discharges holds in other gases
as well.

We also investigate the consequences of pulse repetition for the discharge
morphology. Remaining pre-ionization of previous discharge pulses
creates thicker streamers and a lower branching rate as shown previously
in~\cite{Nijdam2011c}. We find that the effects of pre-ionization
are constrained to the volume of the previous discharge, in agreement
with estimates of the diffusion length between pulses. Like in the
experiments of Takahashi~\cite{Takahashi2011}, the pre-ionization
has localised effects on the discharge.

\section*{References}

\bibliographystyle{unsrt}
\bibliography{Streamers}

\end{document}